# A Quantum-Inspired Framework for Subjective Evaluation: Cognitive Polarization and Entropic Measures

**Bumned Soodchomshom**

Department of physics, Faculty of science, Kasetsart University, Bangkok 10900, Thailand

E-mail: fscibns@ku.ac.th; bumned@hotmail.com

**Abstract**

We propose a quantum-inspired framework to model subjective evaluation processes using state vectors in Hilbert space. In this approach, individual preferences are represented as cognitive states polarized between "like" and "dislike", enabling a continuous interpretation of evaluative attitudes. The evolution of these states is characterized on the Bloch sphere, and the cognitive coherence is interpreted geometrically. To further analyze the uncertainty and diversity in subjective preferences, we introduce both Shannon entropy (at the individual level) and Von Neumann entropy (at the group level) into the framework. A small-scale simulated dataset is used to conceptually demonstrate how these entropy measures can reveal internal indecisiveness and collective incoherence. The model offers a physically grounded and mathematically expressive tool for quantifying subjectivity.

**Keywords:**
Quantum cognition; Subjective evaluation; Hilbert space; Bloch sphere; Shannon entropy; Von Neumann entropy; Cognitive uncertainty

## 1. Introduction

Human preference and subjective evaluation have traditionally been modeled within the frameworks of classical decision theory and utility maximization [1,2]. These classical models assume that individuals hold stable, pre-existing preferences, which are merely revealed through judgment. However, empirical evidence has consistently demonstrated that human evaluations are inherently unstable, context-sensitive, and temporally dynamic [3,4]. Preferences may shift with framing, fluctuate over time, and remain indeterminate until triggered by external stimuli. To address these anomalies, the field of quantum cognition has emerged, applying mathematical structures from quantum mechanics—particularly Hilbert space formalism and probabilistic amplitudes—to model cognitive processes. Seminal works by Busemeyer et al. [5], Bruza et al. [6], and others have shown that core phenomena such as disjunction effects, response interference, and order effects [7] can be naturally captured using quantum probability theory. These models represent mental states as vectors in Hilbert space that evolve under contextual influences and are probabilistically projected onto outcome bases during decision-making. Quantum decision models can adopt either a closed-system Schrödinger dynamics, in which contextual influences are not explicitly accounted for, or an open-system framework, where interaction with the environment naturally incorporates contextual and external factors. Both approaches have been applied in the literature on quantum decision theory, reflecting complementary perspectives on how cognitive states evolve under different conditions. Despite these advances, many existing models still rely on the notion of **projective measurement and state collapse**, which assumes that cognitive states are instantaneously reduced to definite values upon measurement. While suitable for modeling binary choices or discrete outcomes, this assumption limits their applicability to **continuous and reversible evaluative processes**, such as taste, aesthetic judgment, or affective response. These phenomena require a formalism that accounts for **gradual, non-destructive, and context-sensitive state transitions**—features not well captured by collapse-based frameworks.

Recent developments in quantum decision theory (QDT) have begun to address this gap by introducing tools from open quantum systems and weak measurement theory. These include mixed states, non-projective measurements (e.g., POVMs), and continuous-time evolution under non-Hermitian or time-dependent Hamiltonians [8,9]. Notably, Khrennikov [10], Favre et al. [11], and Pothos & Busemeyer [12] propose generalized quantum models that allow for partial state readouts without full collapse, aligning more closely with the dynamic nature of human perception and decision-making. However, **a key limitation remains**: most prior work focuses on modeling choices or belief updates as isolated events rather than continuous evaluative processes. Few models explore how subjective impressions evolve in real-time under ongoing interaction with a stimulus, particularly in the **absence of forced collapse**. Moreover, existing frameworks lack explicit tools to describe and quantify **collective evaluation patterns** across populations.

In this paper, we propose a **quantum-inspired conceptual model** of subjective evaluation that directly addresses these limitations. Our framework models individual cognitive states using the Bloch sphere representation and time-dependent Hamiltonian dynamics. Evaluation is treated not as a discrete measurement, but as a **weak or generalized readout** of a dynamically evolving state vector [16]. This perspective avoids state collapse and accommodates reversibility, graded preference, and temporal fluctuations. We focus on the illustrative case of **taste testing**, where an unfamiliar stimulus induces cognitive dynamics governed by a stimulus-specific potential H(t). The resulting preference state is evaluated through projection along the "like" axis, interpreted as a probabilistic amplitude squared (e.g., mapped to a 0–10 scale). Furthermore, we introduce the concept of **like-polarization**, a population-level metric analogous to spin polarization, which quantifies the collective orientation of preferences across evaluators. In addition to modeling cognitive states geometrically on the Bloch sphere, we propose an entropy-based interpretation to quantify subjective uncertainty and inter-individual diversity. This allows the framework to extend beyond representation into the realm of **information-theoretic quantification** of evaluative behaviors. This addition enables group-level analysis and opens the door to applications in consumer research, satisfaction analytics, and human-machine interaction. The rest of the paper develops the mathematical formalism, simulates dynamic state evolution under various interactions, and discusses implications for behavioral modeling and affective computation.

## 2. Theoretical Model

We model the subjective evaluation process using a two-dimensional Hilbert space, where the cognitive state of the subject is represented by a normalized state vector $\left|\psi(t)\right\rangle$ on the Bloch sphere [13]. The basis states $\left|\text{like}\right\rangle$ and $\left|\text{dislike}\right\rangle$ span the preference space. Before the subject experiences the stimulus (e.g., tastes the food), the initial cognitive state is in a neutral superposition:

$$\left|\psi(0)\right\rangle = \alpha_0\left|\text{like}\right\rangle + \beta_0\left|\text{dislike}\right\rangle, \tag{1}$$

where $\left|\text{like}\right\rangle = \begin{pmatrix}1\\0\end{pmatrix}$ and $\left|\text{dislike}\right\rangle = \begin{pmatrix}0\\1\end{pmatrix}$ . $\alpha_0$ and $\beta_0$ are initial amplitude for $\left|\text{like}\right\rangle$ and $\left|\text{dislike}\right\rangle$, respectively. The stimulus introduces a time-dependent potential V(t), generating evolution governed by a unitary operator:

$$\left|\psi(t)\right\rangle = U(t)\left|\psi(0)\right\rangle = \mathcal{T}\exp\left(-\frac{i}{\hbar_{eff}}\int_0^t H(\tau)\,d\tau\right)\left|\psi(0)\right\rangle, \tag{2}$$

where, $\hbar_{eff}$ stands for an effective Planck constant, which serves only as a scaling parameter for time–energy units. Following prior works in quantum-like decision modeling, $\hbar_{eff}$ can be set equal to unity without loss of generality, and we clarify in the text that it has no physical interpretation in human cognition. $H(t) = H_0 + V(t)$ is the effective Hamiltonian representing the affective and sensory interaction with the stimulus, and $\mathcal{T}$ is the time-ordering operator [14]. Here, $H(t) \approx V(t)$, when $H_0 = 0$.

At any time t, the **subjective score** s(t)∈[0,1] is modeled as a **weak measurement outcome**, approximately given by:

$$s(t) \approx \left|\langle \text{like} \mid \psi(t)\rangle\right|^2 = \alpha(t)^2. \tag{3}$$

Here, $s(t)$ denotes the probability—estimated via a weak, non-projective readout—of the **'like'** outcome at time t. Importantly, Eqn (3) should be understood as an approximate weak readout rather than a full projective collapse. In this sense, the expression reflects a continuous projection process: repeated weak samplings during the short-term sensory memory window yield a time-averaged value of $\alpha(t)^2$, which represents the reported score while leaving the underlying superposition largely coherent. This interpretation treats the score not as a collapse to a binary decision, but as a continuous projection reflecting the current cognitive disposition [15,16].

### 2.1 Cognitive State Representation on the Bloch Sphere

In our model, the subjective cognitive state of an individual is treated as a pure quantum-like state in a two-level Hilbert space. This state can be uniquely represented by a point on the surface of the Bloch sphere, where the north pole corresponds to the basis state $\left|\text{like}\right\rangle$, and the south pole corresponds to $\left|\text{dislike}\right\rangle$. Any pure state $\left|\psi(t)\right\rangle$ of the system can thus be written in the form:

$$|\psi(t)\rangle = \cos\left(\frac{\theta(t)}{2}\right)|\,\text{like}\rangle + e^{i\phi(t)}\sin\left(\frac{\theta(t)}{2}\right)|\,\text{dislike}\rangle\,. \quad (4)$$

Here, $\theta(t)$ and $\phi(t)$ are the polar and azimuthal angles on the Bloch sphere as depicted in Fig.1, respectively, encoding the evolving cognitive disposition toward the stimulus [17].

### 2.2 Effect of External Stimuli via Hamiltonian Dynamics

The interaction between the subject and an external stimulus (e.g., tasting unfamiliar food) is modeled through a time-dependent Hamiltonian $H(t) = V(t)$, which governs the unitary evolution of the cognitive state $\left|\psi(t)\right\rangle$ according to the Schrödinger equation:

$$i\hbar_{eff}\frac{d}{dt}|\psi(t)\rangle = H(t)\,|\psi(t)\rangle. \quad (5)$$

As illustrated in **Fig. 2**, this cognitive dynamics can be interpreted as a physical analogy to quantum evolution. The brain initially generates a subjective preference state $\left|\psi(t_0)\right\rangle$ in the form of a coherent superposition between the basis states |like⟩ and |dislike⟩ (Fig. 2a). Here, "coherent" means that the relative phase between the like and dislike components is preserved, allowing interference between them, in contrast to an incoherent mixture where phase information is lost. Upon encountering the stimulus (e.g., food), the sensory system performs a biosensing process (Fig. 2b), converting chemical interaction into electric signals. These signals induce a time-dependent internal modulation that mimics a Hamiltonian H(t), which acts on the state vector and drives the cognitive dynamics over time.

We express the Hamiltonian in terms of Pauli matrices $\vec{\sigma}=(\sigma_x,\sigma_y,\sigma_z)$, [18] as:

$$H(t)=\hbar_{eff}\,\Omega(t)\vec{n}\cdot\vec{\sigma}=\hbar_{eff}\,\Omega(t)[n_x\sigma_x+n_y\sigma_y+n_z\sigma_z]\ . \quad (6)$$

Here $\Omega(t)$ represents the angular frequency of rotation (stimulus strength), $\vec{n}=(n_x,n_y,n_z)$ is a unit vector specifying the axis of rotation in Bloch space. $\vec{\sigma}$ are the standard Pauli matrices: $\sigma_x=\begin{pmatrix}0&1\\1&0\end{pmatrix},\ \ \sigma_y=\begin{pmatrix}0&-i\\i&0\end{pmatrix},\ \ \sigma_z=\begin{pmatrix}1&0\\0&-1\end{pmatrix}$. The formal solution to this time evolution is given by the unitary operator $|\psi(t)\rangle=U(t,t_0)|\psi(t_0)\rangle$. If the Hamiltonian is constant during the interaction period (impulsive approximation), the operator simplifies to:

$$U(t,t_0)=\exp\left[-i\Theta(t)\vec{n}\cdot\vec{\sigma}\right]=cos\Theta(t)\cdot I-isin\Theta(t)\cdot(\vec{n}\cdot\vec{\sigma}),$$

with: $$\vec{n}\cdot\vec{\sigma}=n_x\sigma_x+n_y\sigma_y+n_z\sigma_z=\begin{pmatrix}n_z & n_x-in_y\\ n_x+in_y & -n_z\end{pmatrix}, \quad (7)$$

where $\Theta(t)=\int_{t_0}^{t}\Omega(\tau)\,d\tau$. Here, the effective Planck constant $\hbar_{\text{eff}}$ cancels between the Hamiltonian in Eqn (6) and the denominator of the evolution operator in Eqn (2), so it no longer appears in Eqn (7). The unit vector $\vec{n}$ is fixed, while the time dependence is fully absorbed into the rotation angle Θ(t). This rotation operator causes the state vector on the Bloch sphere to process around the axis $\vec{n}$, producing an updated state $|\psi(t)\rangle$. As shown in Fig. 2c, this final state $|\psi(t_f)\rangle$ encodes the subject's post-interaction preference orientation—allowing **a weak, continuous-scale measurement** [16] to yield a score (e.g., 0–10) without collapsing the state.

To clarify how the amplitudes can be read out without invoking full collapse, we formalize the readout as a non-projective (POVM) measurement. Let the subjective state be

$$|\psi\rangle=\alpha_f\,|\,\text{like}\rangle+\beta_f\,|\,\text{dislike}\rangle\,.$$

We define Kraus operators, analogous to a real quantum system, as

$$M_+=\sqrt{1-\epsilon}\,|\,\text{like}\rangle\langle\text{like}\,|+\sqrt{\epsilon}\,|\,\text{dislike}\rangle\langle\text{dislike}\,|,$$
$$M_-=\sqrt{\epsilon}\,|\,\text{like}\rangle\langle\text{like}\,|+\sqrt{1-\epsilon}\,|\,\text{dislike}\rangle\langle\text{dislike}\,|,$$

where $0<\epsilon\ll 1$, so that $M_+^{\dagger}M_+ + M_-^{\dagger}M_- = I$. The probability of obtaining outcome “+” is

$$p_+=\langle\psi\,|\,M_+^{\dagger}M_+\,|\,\psi\rangle=(1-\epsilon)\,|\,\alpha_f\,|^2+\epsilon\,|\,\beta_f\,|^2,$$

and the corresponding post-measurement state, which is not collapsed, is

$$|\psi_{\text{post},+}\rangle = \frac{M_+ |\psi\rangle}{\sqrt{p_+}}.$$

Because $0 < \epsilon \ll 1$, the disturbance to the superposition is small ($\sim O(\epsilon^2)$), and the cognitive state retains coherence even after reporting a score. Repeated weak samplings within the short temporal window in which the gustatory trace is sustained allow the brain to average over multiple partial readouts. The reported liking level thus approximates the amplitude squared,

$$L \approx |\alpha_f|^2,$$

without requiring collapse to a binary choice. Neurocognitively, this corresponds to sustained activity in gustatory and prefrontal circuits, which enables repeated internal access to the same sensory trace—functionally analogous to a **continuous weak measurement** [16].

## 3. Model-Based Simulation of Evaluation as Quantum State Evolution

To illustrate the operational mechanism of our model, we consider a subjective evaluation process—such as tasting a food item—as a quantum-like state transition on the Bloch sphere. The cognitive state of the evaluator is initially prepared in a superposition of the two basis states: $|\text{like}\rangle$ and $|\text{dislike}\rangle$, which span the two-dimensional Hilbert space of evaluative preference.

At time $t_0$, the subject has not yet interacted with the stimulus. Their cognitive state is denoted by $|\psi(0)\rangle = \alpha_0 |\text{like}\rangle + \beta_0 |\text{dislike}\rangle$, where the coefficients $\alpha_0, \beta_0 \in \mathbb{C}$, and $|\alpha_0|^2 + |\beta_0|^2 = 1$. This state can be visualized as a vector on the Bloch sphere, parameterized by angles $\theta_0$ and $\phi_0$, as shown in **Fig. 3b** (left) [13,14]. Upon exposure to the stimulus, a time-dependent potential $V(t)$ becomes active during a finite interval (see **Fig. 3a**). This interaction induces a Hamiltonian $H(t) = \hbar\Omega(t)\vec{n}\cdot\vec{\sigma}$, where $\vec{n}$ is a unit vector indicating the axis of cognitive rotation, and $\Omega(t)$ quantifies the intensity of the stimulus-induced effect. This generates a unitary time evolution [14,18] as

$$|\psi(t_f)\rangle = U(t_f, t_0)|\psi(t_0)\rangle.$$

This evolution manifests as a rotation of the cognitive Bloch vector (Fig. 3b, right), modifying the internal disposition toward "like" or "dislike" based on the character and strength of the stimulus [17]. For example, a pleasant taste might rotate the state closer to the north pole (increasing $|\alpha_f|^2$ ), while an unpleasant stimulus might shift it toward the south pole.

**Importantly**, the perceived interaction $V(t)$ is not universal across all participants. It depends on individual differences in sensory sensitivity, cognitive bias, expectation, and psychological state. Thus, while the stimulus (e.g., the same food sample) is identical, its cognitive representation $V_i(t)$ may vary per individual "$i$". In some cases, if the participant exhibits no change in preference magnitude—i.e., the final $\theta$ remains equal to the initial $\theta_0$—we may interpret $V(t)$ as having only induced a rotation around the z-axis, i.e., modifying $\phi$ without changing $\theta$. This corresponds to a **pure phase rotation**, preserving the "degree of liking" while altering internal contextual orientation.

**Fig. 3c** illustrates two cases: an **unbiased evaluator** whose initial state lies on the equator ($|\alpha_0|^2 = |\beta_0|^2$), and a **biased evaluator** whose state is pre-tilted due to prior preference ($|\alpha_0|^2 \neq |\beta_0|^2$). In both cases, the stimulus interaction alters the amplitudes to $\alpha_f, \beta_f$, which encode the final judgment. The output of the evaluation is a subjective score, given by the squared projection onto the "like" axis:

$$\text{score} = |\langle \text{like} | \psi(t_f) \rangle|^2 = |\alpha_f|^2$$

This value lies in the range of [0,1], allowing for continuous-scale assessment (e.g., 0–10), and it **does not collapse the state**. The evaluator's mental state remains coherent, enabling possible re-evaluation or temporal fluctuation [15,16]. In this framework, the evaluation process is conceptualized not as a binary decision, but as a **weak measurement of a dynamically evolving mental state**, guided by physical-like interaction principles. The use of Bloch sphere dynamics provides a transparent geometric interpretation of how stimulus quality and duration influence cognitive preference. It also accommodates heterogeneity among participants, making it suitable for modeling subjective perception in applications such as consumer testing, neuroeconomics, and psychometric diagnostics.

## 4. Discussion

This work proposes a quantum-theoretical framework for modeling subjective evaluation—such as tasting a novel food—by representing the internal cognitive state of the evaluator as a quantum state evolving over time on the Bloch sphere. In contrast to classical models that assume pre-defined preferences and discrete evaluations, our model frames the decision-making process as a continuous evolution of a cognitive vector under the influence of an external stimulus, modeled as a time-dependent Hamiltonian $H(t) \approx V(t)$. A key theoretical contribution of this work is the explicit use of the Bloch sphere to represent the dynamic evolution of subjective states, where the angle $\theta(t)$ encodes the intensity of preference (from complete dislike to complete like), and $\phi(t)$ may capture contextual or affective bias that shifts without altering the fundamental preference axis. This geometrical interpretation extends the utility of quantum cognitive models beyond binary decisions and enables a clear visual and analytical representation of dynamic preference modulation.

The role of $V(t)$—the interaction potential—as a context-specific, person-dependent operator reflects the inherently individualized nature of human perception. Different participants may experience different effective Hamiltonians based on sensory sensitivity, emotional state, memory, or prior experience, even when exposed to the same stimulus. This introduces a flexible yet mathematically consistent structure for encoding inter-individual variability in subjective response. Importantly, the act of assigning a score (e.g., 0–10) is interpreted not as a projective measurement (which collapses the state) but rather as a weak or generalized measurement (e.g.,

POVM), as discussed in section2.2, which reveals partial information about the current state without destroying its coherence [19]. This distinction allows the model to capture phenomena such as intra-individual fluctuation of ratings over time and the persistence of mixed feelings post-evaluation—both of which are difficult to reconcile with collapse-based models.

When extending this framework to a population of evaluators, it becomes possible to define a statistical measure akin to spin polarization in quantum mechanics. For each individual, their cognitive state is represented as a Bloch vector with a projection on the "like" axis. By averaging these projections across all evaluators, we define a collective measure—termed **like-polarization** *($P_{\text{like}}$)*—that captures the population-level tendency toward liking or disliking a stimulus. This can be expressed mathematically as:

$$P_{\text{like}} = \frac{1}{N}\sum_{i=1}^{N}\left\langle \psi_i(t_f)\middle|\sigma_z\middle|\psi_i(t_f)\right\rangle = \langle\alpha_f^2\rangle - \langle\beta_f^2\rangle = 2\langle\alpha_f^2\rangle - 1, \quad (8)$$

where $\langle\alpha_f^2\rangle$ is the average probability amplitude squared of the "like" component for each evaluator. The value of $P_{\text{like}}$ ranges from -1 (universal dislike) to +1 (universal like), with 0 indicating a neutral or balanced collective response. **As illustrated in Fig. 4,** when a population of N evaluators interacts with a stimulus, their cognitive states evolve into distinct outcome groups: those who prefer the stimulus (**like group**, where $\alpha_f^2 > \beta_f^2$, those who express no clear preference (**neutral group**, with $\alpha_f^2 = \beta_f^2$, and those who dislike the stimulus (**dislike group**, where $\alpha_f^2 < \beta_f^2$. These categories reflect the projection of each individual's state onto the “like” axis after interaction. The like-polarization value $P_{\text{like}}$ summarizes the overall directionality of preference in the group: $P_{\text{like}} = 1$ indicates unanimous liking, $P_{\text{like}} = -1$ reflects unanimous disliking, and $P_{\text{like}} = 0$ implies a balanced or ambivalent collective response. Fig. 4 thus visualizes how quantum cognitive dynamics manifest at the ensemble level. Here, the ensemble level refers to the aggregation of cognitive states across individuals. The polarization index $P_{\text{like}}$ not only reflects the group’s mean tendency but also the dispersion of evaluative orientations. This provides a dynamic metric of consensus or disagreement, extending prior quantum-like models focused mainly on individual dynamics. This measure serves as a dynamic, time-dependent summary of evaluative consensus, reflecting both the mean orientation and the dispersion of cognitive states within a group. It is especially relevant in applications such as market research or sentiment analysis, where aggregated preference data are critical.

The implications of this framework extend to various domains of behavioral science and cognitive modeling. While several quantum-inspired cognitive models have addressed context dependence and order effects [5–7,9], few provide a continuous-time formalism for subjective evaluation. For instance, Busemeyer et al. [5] focus on decision interference in discrete trials, and Khrennikov [10] models contextuality without direct temporal evolution. Our model extends these approaches by incorporating explicit Bloch sphere trajectories and Hamiltonian dynamics, enabling a fine-grained description of evolving preference strength. This positions our work within the growing effort to bridge quantum formalism with psychologically grounded dynamics [12, 14, 16]. In consumer research, for instance, it offers a formal explanation for varying customer satisfaction ratings in response to identical products. In affective computing and human–machine interaction, it provides a basis for designing interfaces that interpret gradual shifts in user sentiment. Additionally, the model aligns with current advances in quantum-like modeling of cognition, where probabilistic and dynamical properties of cognitive processes are treated in line with open quantum systems. Nonetheless, several limitations remain. Furthermore, the current model assumes ideal unitary dynamics without decoherence, which may not be realistic in cognitive systems where noise and attention fluctuations play a significant role. Models such as Wiseman and Milburn's [16] quantum control frameworks, or open quantum system approaches from Scully and Zubairy [17], suggest that environment-induced decoherence can substantially alter state trajectories. Incorporating these effects into future versions of our model would enhance its predictive validity and biological plausibility. This model has not yet been empirically tested, and the specific form of $V(t)$ remains to be derived from physiological or psychological data. However, a simple model for $V(t)$ may be easily introduced as a step function, given by

$$V(t)=\begin{cases} 0 & t\le t_0 \\ \hbar\Omega\sigma_y & t_0<t<t_f \\ 0 & t\ge t_f \end{cases}, \tag{9}$$

for unbiased case $|\psi(t_0)\rangle=\frac{1}{\sqrt{2}}|\text{like}\rangle+\frac{1}{\sqrt{2}}|\text{dislike}\rangle$. This V(t) leads to the time evolution operator of the form

$$U(t_f,t_0)=\begin{cases} I & t\le t_0 \\ e^{-i\Omega\sigma_y(t-t_0)} & t_0<t<t_f \\ e^{-i\Omega\sigma_y(t_f-t_0)} & t\ge t_f \end{cases}. \tag{10}$$

By setting $t_0=0$, we get $|\psi(t)\rangle=U(t,0)|\psi(0)\rangle=\alpha(t)|like\rangle+\beta(t)|dislike\rangle$, where

$$\alpha(t_f\ge t\ge 0)=\frac{1}{\sqrt{2}}\cos(\Omega t)-\frac{1}{\sqrt{2}}\sin(\Omega t),$$

and $$\beta(t_f\ge t\ge 0)=\frac{1}{\sqrt{2}}\sin(\Omega t)+\frac{1}{\sqrt{2}}\cos(\Omega t).$$

As illustrated in **Fig. 5,** the behavior of $\alpha(t)^2$—the probability of the "like" component—varies over time depending on the sign and magnitude of the stimulus strength parameter for $|\Omega|<\frac{\pi}{2t_f}$ in the Hamiltonian $H(t)=V(t)$ defined in Eq.9. When $\Omega=0$, the cognitive state remains unchanged (neutral evaluation), and $\alpha(t)^2$ stays constant. For $\Omega<0$, the amplitude $\alpha(t)^2$ increases over time, representing a growing preference toward "like." Conversely, for $\Omega > 0$, the amplitude decreases, indicating a shift toward "dislike." This simulation confirms that the model captures individualized, context-sensitive evaluation dynamics through simple Hamiltonian rotation, and provides a basis for analyzing subjective responses using a unified quantum framework.

In addition to the geometrical representation of cognitive states on the Bloch sphere, we conducted an information-theoretic analysis of evaluative uncertainty using both Shannon and Von Neumann entropy. For each participant "*i*", the preference score $\alpha_{f,i}^2 \equiv p_i \in [0,1]$ was interpreted as the probability of "liking" the coffee. The classical **Shannon entropy** for that individual is given by

$$H(p_i) = -p_i \log_2 p_i - (1-p_i)\log_2(1-p_i) , \quad (11)$$

which was originally introduced in the context of communication theory [20]. This entropy quantifies the degree of uncertainty in their judgment: values near 0 correspond to strong certainty (either like or dislike), while values near 1 indicate maximum ambiguity. The **average Shannon entropy** over all participants provides a global measure of decisiveness within the population

$$\bar{H} = \frac{1}{N}\sum_{i=1}^{N} H(p_i) .$$

To capture the structure of collective evaluative diversity, we modeled each individual's preference as a pure state on the Bloch sphere:

$$|\psi_i\rangle = \sqrt{p_i}\,|\,\text{like}\rangle + e^{i\phi_i}\sqrt{1-p_i}\,|\,\text{dislike}\rangle . \quad (12)$$

We then constructed the **density matrix** for the ensemble as:

$$\rho = \frac{1}{N}\sum_{i=1}^{N} |\psi_i\rangle\langle\psi_i| .$$

The **Von Neumann entropy**, defined as

$$S(\rho) = -\mathrm{Tr}(\rho \log_2 \rho) , \quad (13)$$

which is a fundamental measure in quantum information theory [21]. This captures the degree of **mixedness** of the collective state—i.e., the cognitive diversity or incoherence arising from angular dispersion among pure-state preferences. Here, angular dispersion on the Bloch sphere represents the heterogeneity of evaluative orientations across individuals. Such dispersion increases the mixedness of the ensemble state, which is precisely what the Von Neumann entropy quantifies as cognitive diversity. Notably, even though each individual is modeled as occupying a pure state, the ensemble state $\rho$ is generally mixed unless all $|\psi_i\rangle$ are identical. This reveals how cognitive diversity at the group level can be formally quantified via quantum entropy. To conceptually illustrate the proposed entropy-based interpretation, we simulated a small-scale dataset $p_i = \{0.1, 0.3, 0.7, 0.9, 0.5, 0.95\}$ reflecting plausible user preferences. The calculated average Shannon entropy was $\bar{H} \approx 0.66$, and the Von Neumann entropy of the ensemble state was $S(\rho) \approx 0.53$. These values suggest a moderate degree of individual hesitation and cognitive diversity within the simulated population. Together, these two entropy measures offer a layered view of evaluative uncertainty: **Shannon entropy** reflects intra-individual ambiguity, while **Von Neumann entropy** captures inter-individual cognitive dispersion [22].

These results also suggest a deeper interpretation of the cognitive evaluator—not as a deterministic agent, but as a quantum-like system whose internal state exists in superposition and evolves continuously under contextual interaction. Within this perspective, one may view human evaluators as "*natural cognitive qubits*" whose judgments emerge not from binary choice, but from time-dependent amplitude distributions shaped by interaction potentials. The use of Bloch sphere dynamics and weak measurement theory thus supports the notion that human preference operates more like a quantum system than a classical processor. Future work could explore integrating neural or physiological correlates (e.g., EEG-based preference markers) as empirical proxies for state vector orientation, drawing inspiration from quantum neuroeconomics [9,23]. Additionally, stochastic Schrödinger equations [14, 16] may offer a useful extension to model noisy, real-world evaluative processes. This perspective opens new directions for exploring biological cognition as a substrate for quantum-like computation and perception [19,23].

Several prior studies have explored quantum-like frameworks for cognition and decision-making. Brody [24] formalized cognitive dynamics via unitary evolution of a qubit-like state with collapse governed by the Born rule, thereby interpreting squared amplitudes as probabilities for binary outcomes. Kurtev [25] emphasized wave-like patterns in parameter space as evidence of macroscopic quantum-like effects, without specifying a measurement framework. Broekaert et al. [26] provided a broader review of dynamical options—including Schrödinger evolution, decoherence, and open system approaches—highlighting the flexibility of quantum formalism in cognitive modeling. Khrennikov and Asano [27] modeled brain information processing through open quantum systems theory, where decisions emerge via decoherence without invoking explicit collapse. Similarly, Asano et al. [28] described decision-making as mental-state decoherence driven by interaction with memory, and Asano et al. [29] introduced lifting channels to capture non-Markovian fluctuations in game-theoretic contexts.

Our proposal diverges from these approaches in two critical respects. First, we avoid both full collapse and complete decoherence. Instead, we formalize subjective evaluation through a non-projective measurement (POVM) scheme, where Kraus operators induce only weak disturbance to the cognitive state. This permits the brain to access partial information about the amplitudes while preserving coherence. Second, rather than treating decision outcomes as binary, we interpret the squared amplitudes directly as degrees of liking or disliking. Continuous weak readouts, enabled by short-term memory traces of sensory experience (e.g., gustatory cortex activity after tasting coffee), provide a mechanism for averaging across multiple partial samplings. In this way, the reported subjective score approximates the underlying amplitudes, distinguishing our model from previous decoherence-based or game-theoretic formulations. Thus, while prior quantum-like models emphasize collapse or decoherence as mechanisms for decision emergence, our framework introduces a weak continuous measurement perspective that allows subjective evaluations to be reported without destroying the coherence of the mental state. This difference underlines the novelty of our approach relative to the existing literature.

## 5. Conclusion

This work presents a conceptual, quantum-inspired framework for modeling subjective evaluation as a dynamic cognitive process on the Bloch sphere. By introducing a time-dependent Hamiltonian to describe how a stimulus—such as tasting unfamiliar food—modulates cognitive state evolution, the model departs from traditional static or binary representations of preference. Evaluation is treated as a weak or generalized measurement, capturing partial information without collapsing the mental state. This preserves the coherence of cognitive superpositions and enables temporal reversibility in judgment. A key contribution is the introduction of **like-polarization**, a population-level metric analogous to spin polarization, which quantifies collective evaluative tendencies across individuals. This construct bridges individual-level cognitive dynamics with group-level statistical analysis, offering a new tool for quantifying subjective trends in domains such as consumer research, behavioral science, and psychometrics. Importantly, this study is primarily theoretical, and while it does not employ empirical experimentation, it includes a conceptual simulation to illustrate the proposed framework. Instead, it provides a mathematically grounded and physically motivated structure that invites future validation and application. By framing evaluation as a continuous quantum-like process rather than a discrete, irreversible choice, this model opens new directions for understanding context-sensitive and temporally evolving human judgments.

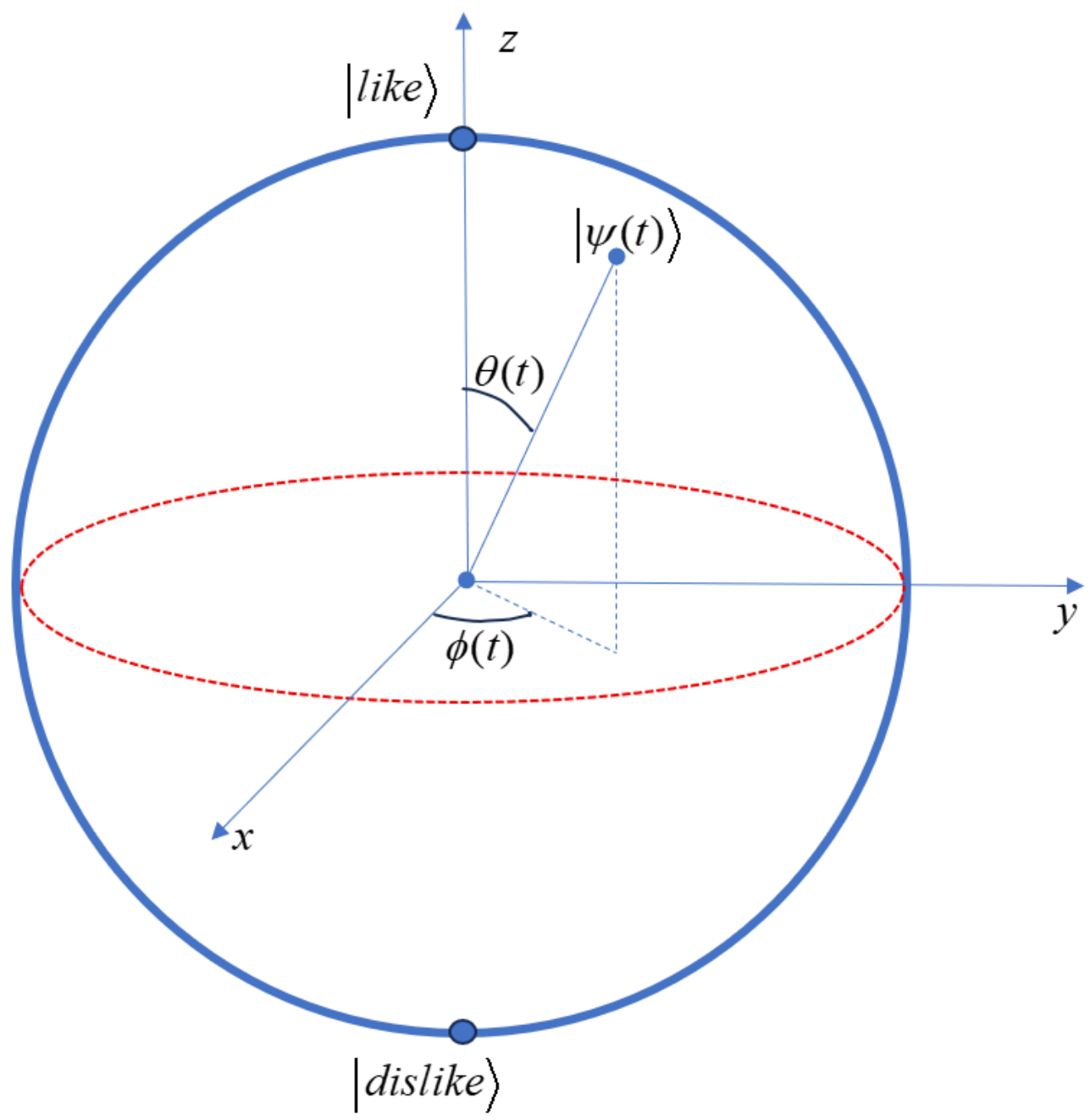

**Figure 1** Representation of the cognitive state vector $|\psi(t)\rangle$ on the Bloch sphere. The basis states $|\text{like}\rangle$ and $|\text{dislike}\rangle$ define the poles along the z-axis. The state $|\psi(t)\rangle$ is parameterized by polar angle $\theta(t)$ and azimuthal angle $\phi(t)$, and evolves under the influence of an external stimulus. The red dashed circle illustrates a possible trajectory of $|\psi(t)\rangle$ due to Hamiltonian-induced rotation around a fixed axis in the Bloch sphere.

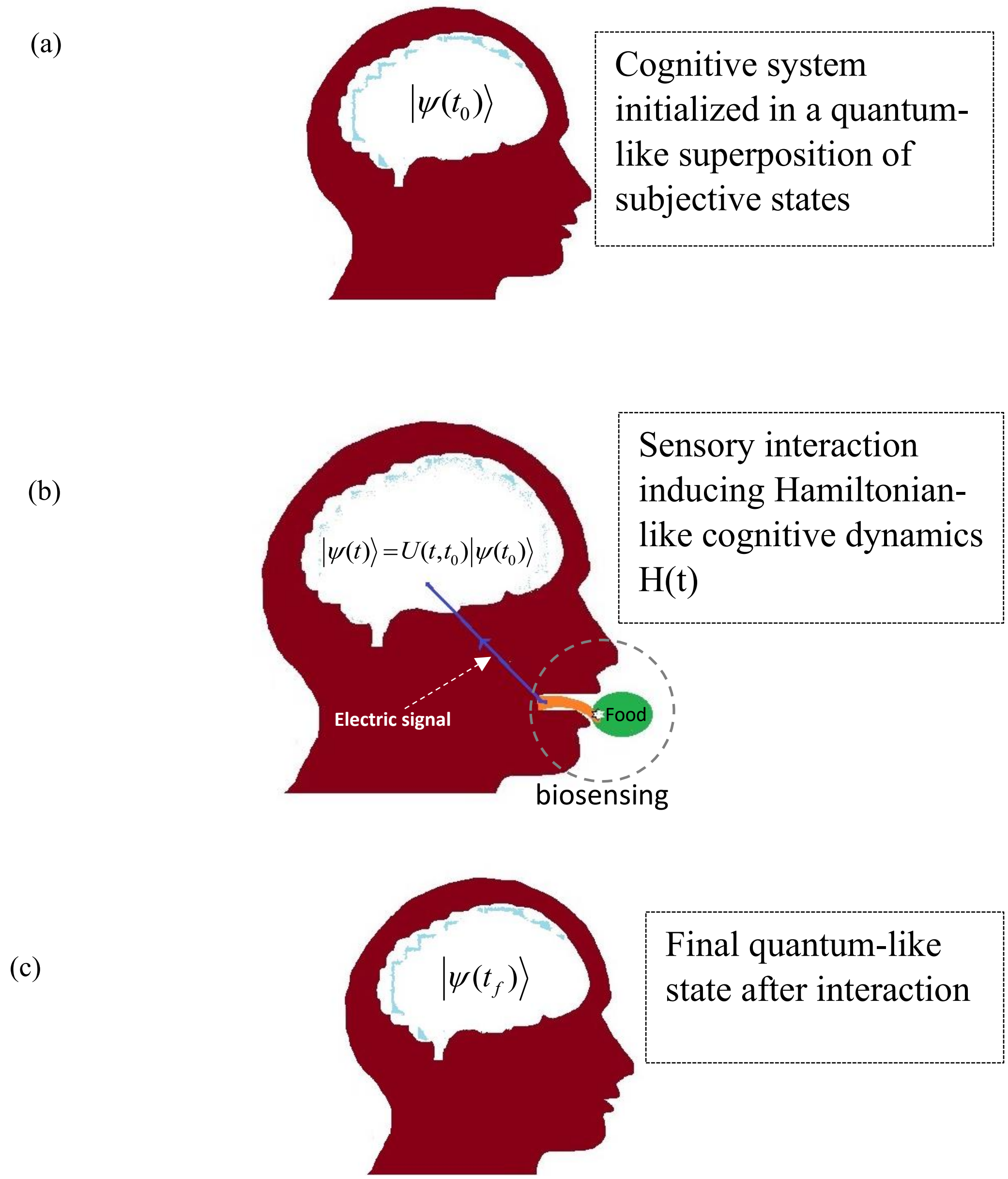

**Figure 2** Schematic representation of the cognitive quantum-like evaluation process during sensory interaction. (a) The initial mental state $|\psi(t_0)\rangle$ is formed prior to stimulus exposure, representing a coherent superposition of subjective preference (e.g., like/dislike). (b) During evaluation, the sensory input (e.g., food) is transduced through a biosensing process, generating electrical signals that act as an effective interaction Hamiltonian H(t), driving the time evolution $|\psi(t)\rangle = U(t,t_0)|\psi(t_0)\rangle$ .(c) After the interaction, the brain reaches a final quantum-like state $|\psi(t_f)\rangle$, which reflects the altered evaluative disposition while retaining coherence—enabling weak or partial measurement (e.g., a continuous score).

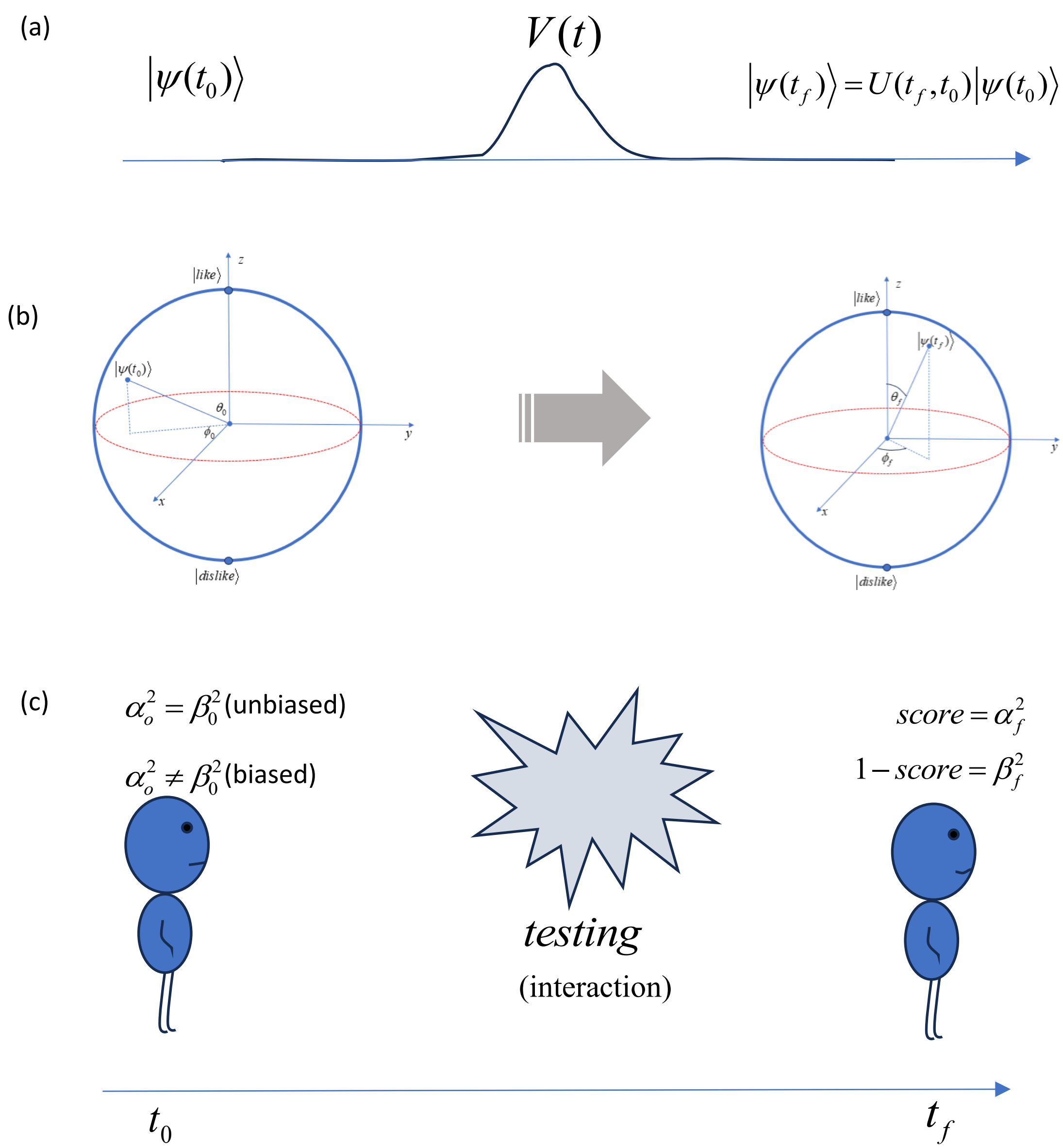

**Figure 3** Illustration of the subjective evaluation process modeled as a time-evolving quantum state under interaction. (a) At the initial time $t_0$, the cognitive state $|\psi(t_0)\rangle$ is represented as a vector on the Bloch sphere. The interaction with a stimulus (e.g., tasting a product) is modeled by a time-dependent potential $V(t)$, which induces unitary evolution $U(t_f, t_0)$ of the state without collapse.(b) This evolution corresponds to a rotation of the Bloch vector, altering the cognitive state's position from $|\psi(t_0)\rangle$ to $|\psi(t_f)\rangle$. (c) Two types of initial conditions are shown: an unbiased evaluator with equal preference amplitudes $\alpha_0^2 = \beta_0^2$, and a biased one where $\alpha_0^2 \neq \beta_0^2$. After interaction, the evaluator expresses a score based on the squared amplitude of the final "like" component, $score = \alpha_0^2$, interpreted as a weak measurement of the evolved state.

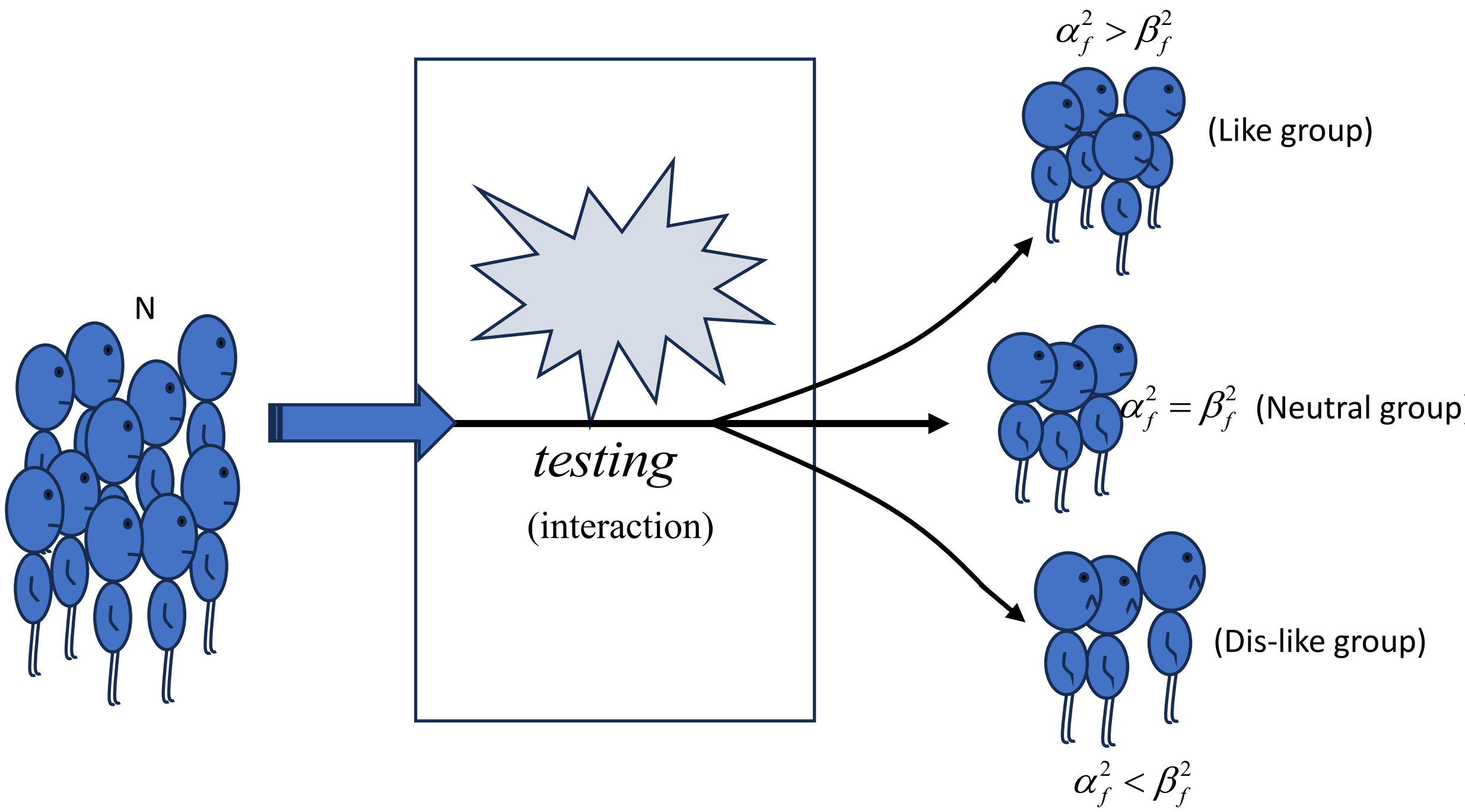

**Figure 4** Illustration of population-level outcomes after quantum-inspired evaluation. A group of N participants undergoes interaction with a stimulus (e.g., taste testing), modeled as a cognitive transition governed by a time-dependent Hamiltonian. Following the evaluation, individual cognitive states evolve into three possible categories based on the final amplitude components: (i) the Like group with $\alpha_f^2 > \beta_f^2$, (ii) the Neutral group where $\alpha_f^2 = \beta_f^2$, and (iii) the Dislike group with $\alpha_f^2 < \beta_f^2$. Aggregating these outcomes enables the computation of the like-polarization metric, defined analogously to spin polarization. A value of +1 indicates unanimous liking across the population, 0 reflects no net preference, and −1 corresponds to unanimous disliking.

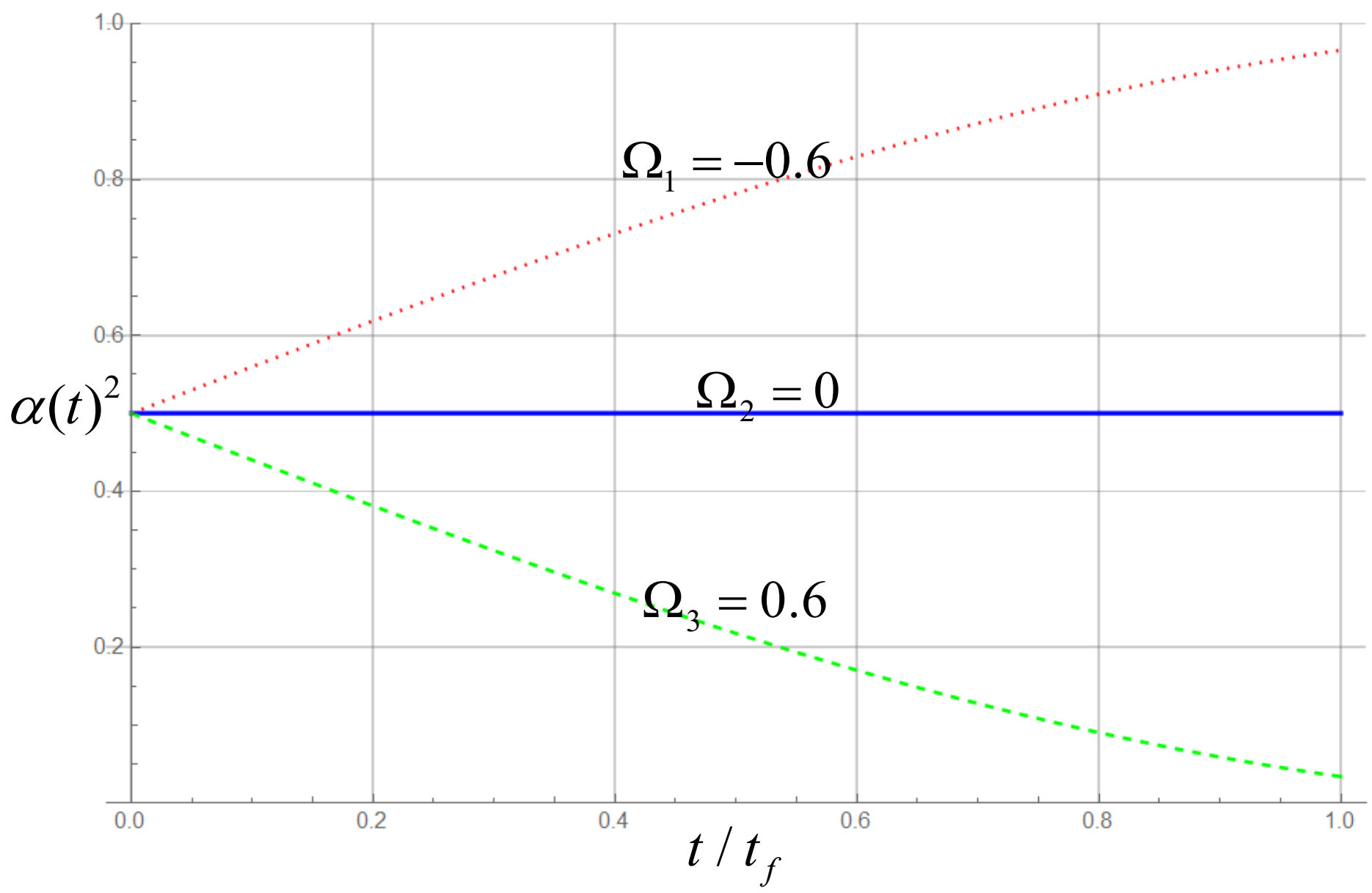

**Figure 5** Time evolution of the squared amplitude $\alpha(t)^2$, representing the probability of the "like" component in the cognitive state under different interaction strengths. The interaction is modeled as a time-independent Hamiltonian $V(t) = \hbar\Omega\sigma_y$, where $\Omega$ quantifies the strength and direction of the stimulus effect. The plot shows three cases: $\Omega = 0$ (solid blue line) corresponds to a neutral response with no cognitive rotation; $\Omega = -0.6$ (dotted red line) induces an increasing preference toward "like"; and $\Omega = 0.6$ (dashed green line) corresponds to a declining preference, interpreted as a shift toward "dislike." This simulation illustrates how variations in $\Omega = 0$ capture individualized evaluator responses under the same stimulus condition.